\documentstyle[sprocl,psfig]{article}

\def\Journal#1#2#3#4{{#1} {\bf #2}, #3 (#4)}

\def\NPB{{\em Nucl. Phys.} B}
\def\PLB{{\em Phys. Lett.}  B}
\def\PRL{\em Phys. Rev. Lett.}
\def\PRD{{\em Phys. Rev.} D}
\def\RMP{\em Rev. Mod. Phys.}

\def\be{\begin{equation}}
\def\ee{\end{equation}}
\def\bea{\begin{eqnarray}}
\def\eea{\end{eqnarray}}

\newcommand{\gsim}{\mathrel{\lower3pt\hbox{$\sim$}}
\hskip-10.5pt\raise1.4pt\hbox{$>$}\;}

\newcommand{\agt}{\mathrel{\lower3pt\hbox{$\sim$}}
\hskip-11.0pt\raise1.5pt\hbox{$>$}\;}

\begin{document}
\begin{flushright}
MADPH--97--1009 \\
July 1997
\end{flushright}
\title{ THE MUON ELECTRIC DIPOLE MOMENT\footnote{
Talk presented by C. Kao 
at the 2nd International Conference on B Physics and CP Violation, 
Honolulu, Hawaii, USA, March 1997. } }
\author{ Vernon Barger, Chung Kao }
\address{Department of Physics, University of Wisconsin, \\
Madison, WI 53706, USA}
\author{ Ashok Das }
\address{Department of Physics and Astronomy, University of Rochester, \\
Rochester, NY 14627, USA}


\maketitle\abstracts{
The electric dipole moment of the muon ($d_\mu$) 
is evaluated in a two Higgs doublet model 
with a softly broken discrete symmetry. 
For $\tan\beta \equiv |v_2|/|v_1| \sim 1$, 
contributions from two loop diagrams involving the $t$ quark 
and the $W$ boson dominate;
while for $\tan\beta \gsim 10$, 
contributions from two loop diagrams involving the $b$ quark 
and the $\tau$ lepton are dominant.
For $8 \gsim \tan\beta \gsim 4$, significant cancellation occurs 
among the contributions from two loop diagrams 
and the one loop contribution dominates for $\tan\beta \sim 7$.
For $\tan\beta \gsim 15$, the calculated $d_\mu$ 
can be close to the reach of a recently proposed experiment 
at the Brookhaven National Laboratory. }

\section{Introduction}

For a particle with spin $\vec{s}$, 
and an electric dipole moment (EDM) $\vec{d} \propto \vec{s}$, 
the interaction Hamiltonian has the following form:
\begin{equation}
H_I = -\vec{d} \cdot \vec{E}, 
\end{equation}
where $\vec{E}$ is an electric field. 
This Hamiltonian violates both parity (P) invariance 
and time-reversal (T) invariance. 
The invariance of CPT implies that 
the combined transformation of charge conjugation (C) and parity (P) 
is not an exact symmetry. \cite{Barr&Marciano,Bernreuther&Suzuki} 
The EDM of a fermion ($d_f$) can be described with an effective Lagrangian 
\begin{eqnarray} 
{\cal L}_{eff} = 
-\frac{i}{2} d_f \bar{\psi} \sigma_{\mu\nu} \gamma_5 \psi F^{\mu\nu}.
\end{eqnarray}
In the non-relativistic limit, the Lagrangian becomes 
${\cal L}_{eff} = - H_I =  d_f  \vec{s} \cdot \vec{E}$.

In the Standard Model (SM) of electroweak interactions, 
CP violation is generated by the Kobayashi-Maskawa (KM) phase.
The electron EDM \cite{EDM1SM} is about 
$8 \times 10^{-41} {\rm e} \cdot {\rm cm}$ 
and the expected muon EDM is about 
$2\times 10^{-38} {\rm e}\cdot {\rm cm}$.
Several experiments have been carried out to search for 
an electron EDM \cite{de} ($d_e$) and a muon EDM \cite{dm} ($d_\mu$). 
At 95\% C.L., the experimental upper limits on $d_e$ and $d_\mu$ are
(i)  $|d_e| <  6.2 \times 10^{-27} {\rm e}\cdot {\rm cm}$, and
(ii) $|d_\mu| < 1.1 \times 10^{-18} {\rm e}\cdot {\rm cm}$.

Recently, a dedicated experiment has been proposed \cite{Yannis} 
to measure the muon electric dipole moment 
at the Brookhaven National Laboratory.
This experiment will be able to improve the measurement of the muon EDM
by at least 4 orders of magnitude.
 
\section{CP Violation from Higgs Exchange}

In multi-Higgs doublet models, a discrete symmetry \cite{Discrete}
is usually required for flavor symmetry to be conserved.
In two Higgs doublet models, this discrete symmetry 
is often chosen to be $\phi_1 \to -\phi_1$ and $\phi_2 \to +\phi_2$. 
If this discrete symmetry is only softly broken \cite{Soft}, 
not only can Higgs boson exchange generate CP violation 
but also flavor changing neutral Higgs interactions 
can be kept at an acceptable level.

A two Higgs doublet model has doublets $\phi_1$ and $\phi_2$. 
After spontaneous symmetry breaking, 
there remain five physical `Higgs bosons': 
a pair of singly charged Higgs bosons ($H^\pm$), 
two neutral CP-even scalars ($H_1$ and $H_2$), 
and a neutral CP-odd pseudoscalar ($A$). 
There are two sources of CP violation in the Higgs potential:
(i)~the mixing of the $A$ with the $H_1$ and the $H_2$, and 
(ii)~the CP violating interaction of $A H^+ H^-$. 

Adopting Weinberg's parameterization \cite{Weinberg} 
we can write the following neutral Higgs exchange propagators as 
\begin{eqnarray}
< H_1 A >_q 
&  = & \frac{1}{2} 
       \sum_n \frac{ \sin 2\beta {\rm Im}Z_{0n} } { q^2-m_n^2 } \nonumber \\
&  = & \frac{1}{2} 
       \sum_n \left[ -\frac{ \cos^2 \beta \cot \beta{\rm Im}\tilde{Z}_{1n} } 
                     { q^2-m_n^2 } 
               +\frac{ \sin^2 \beta \tan \beta{\rm Im}\tilde{Z}_{2n} } 
                     { q^2-m_n^2 } \right] \\
< H_2 A >_q 
& = & \frac{1}{2} 
      \sum_n \frac{ \cos 2\beta {\rm Im}Z_{0n} 
                   -{\rm Im}\tilde{Z}_{0n} } { q^2-m_n^2 }  \nonumber \\
& = & \frac{1}{2} 
      \sum_n \frac{ \cos^2 \beta {\rm Im}\tilde{Z}_{1n} 
                   +\sin^2 \beta {\rm Im}\tilde{Z}_{2n} } { q^2-m_n^2 } 
\label{eq:H&A}
\end{eqnarray}
where the summation is over all the mass eigenstates of neutral Higgs 
bosons. We will assume that the sums are dominated by the lightest 
neutral Higgs boson of mass $m_0$, and drop the sums and indices. 
There are relations among the CP violation parameters: 
${\rm Im} Z_0 +{\rm Im} \tilde{Z}_0 = -\cot^2 \beta {\rm Im} \tilde{Z}_1$, 
and 
${\rm Im} Z_0 -{\rm Im} \tilde{Z}_0 = +\tan^2 \beta {\rm Im} \tilde{Z}_2$.
Employing unitarity constraints, Weinberg has shown that 
$|{\rm Im}Z_1| \le (1/2) |\tan\beta| ( 1+\tan^2\beta)^{1/2}$ and 
$|{\rm Im}Z_2| \le (1/2) |\cot\beta| ( 1+\cot^2\beta)^{1/2}$.

\section{The Muon Electric Dipole Moment}


\begin{figure}
\centering
\leavevmode
\psfig{figure=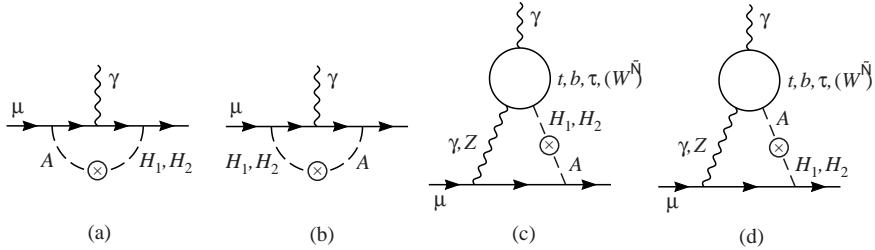,height=1.4in}
\caption{
Feynman diagrams for leading contributions to the muon EDM. 
\label{fig:muon1}}
\end{figure}

The Feynman diagrams contributing to the muon electric dipole moment 
are shown in Figure 1. There are many more diagrams involving the $W$ boson 
that are not shown in this figure.
We consider only two-loop diagrams with an intermediate photon 
since the diagrams involving an intermediate $Z$ boson are highly suppressed 
by the vector part of the $Ze^+e^-$ couplings. 


In a two Higgs doublet model the electron EDM and the muon EDM 
have contributions from two-loop diagrams containing 
the top quark ($t$-loop) \cite{Top}, 
the gauge bosons ($W$-loop) \cite{Gauge}, and
the charged Higgs boson ($H$-loop) \cite{Higgs}. 

In the contribution from the charged Higgs boson,
CP violation comes from both the neutral Higgs transition propagators 
and the trilinear vertex $A H^+H^-$.\footnote{
Contributions from the $H^\pm$ are not expected 
to be significant, 
since recent CLEO measurements on $B(b \to s\gamma)$ 
constrain $m_{H^\pm}$ to be at least $3M_W$ in Model II. \cite{CLEO} }
For $\tan\beta$ larger than about 10, 
there are dominant contributions from the $b$ quark 
and the $\tau$ lepton \cite{Das&Kao}. 

There are several interesting aspects:
(i) The contributions from the $b$ and the $\tau$ loops 
are proportional to (Im $Z_0$ +Im $\tilde{Z}_0$).
(ii) For the the $t$-loop, the coefficient of the Im $\tilde{Z}_0$ 
is much smaller than that of the Im $Z_0$.
(iii) The $W$-loop does not contribute to the Im $\tilde{Z}_0$ term.


The one-loop contribution to the muon EDM ($d_\mu$) is 
\begin{eqnarray}
d_\mu^{1-{\rm loop}}
& = & \frac{ e m \sqrt{2} G_F \tan^2\beta }{ (4\pi)^2 }
      I(\rho)( {\rm Im}Z_0 +{\rm Im}\tilde{Z}_0 ), \nonumber \\
I(\rho) 
& \equiv & \rho \int_0^1 dx \frac{x^2}{ \rho x^2 -x +1 }, 
\end{eqnarray}
where $\rho = m^2/m_0^2$.
In the limit of $m_0 \gg m$ (i.e. $\rho \ll 1$), 
$I(\rho)$ approaches $ -\rho [ \ln(\rho) +3/2 ]$ and
$ d_{(e,\mu)}^{1-{\rm loop}} \propto (m^3/m_0^2) [ \ln(m^2/m_0^2) ] $, 
for $m = m_e, m_\mu$.


\begin{figure}
\centering
\leavevmode
\psfig{figure=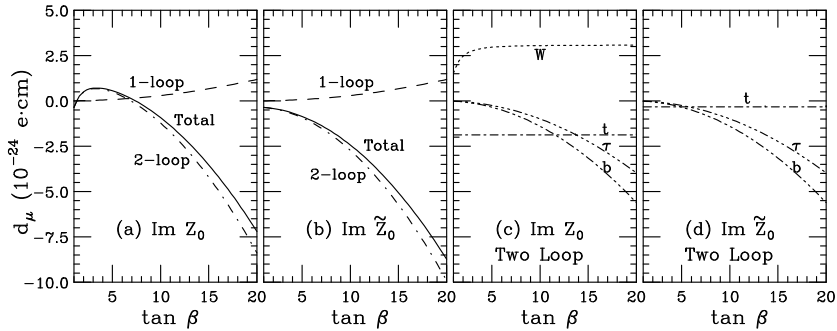,angle=90,height=1.7in}
\caption{
The muon EDM for $m_0 = 100$ GeV in units of Im $Z_0$ and Im $\tilde{Z}_0$. 
\label{fig:muon2}}
\end{figure}

Figure 2 shows the muon EDM generated from one-loop diagrams (dash),
two-loop diagrams (dash-dot) and their total (solid)
in units of (a) Im $Z_0$ and (b) Im $\tilde{Z}_0$. 
Also shown are the contributions from two-loop diagrams
involving the $W$ boson (dot), the $t$ quark (dash-dot),
the $b$ quark (dash-dot-dot), and the $\tau$ lepton (dash-dot-dot-dot),
in units of (c) Im $Z_0$ and (d) Im $\tilde{Z}_0$.
This figure shows the effect of varying $\tan\beta$ on the muon EDM, 
for the case of $m_0 = 100$ GeV.
We note that: 
(i)~for $\tan\beta$ close to one, 
contributions from two loop diagrams involving the top quark 
and the $W$ boson dominate; 
(ii)~for $\tan\beta \agt 10$ contributions 
from two loop diagrams involving the $b$ quark and the $\tau$ lepton 
are dominant; 
(iii)~for $8 \agt \tan\beta \agt 4$, significant cancellation occurs 
among the contributions from two loop diagrams 
and one loop diagrams dominate for $\tan \sim 7$; 
(iv)~for $\tan\beta \sim 1$ or $\tan\beta \agt 10$, 
the muon and the electron EDMs satisfy a simple scaling relation 
$d_\mu \approx (m_\mu/m_e) d_e$;
(v)~for $8 \agt \tan\beta \agt 4$, 
$|d_\mu|$ can be two to three times $|(m_\mu/m_e) d_e|$.

The muon EDM can be expressed as
\begin{eqnarray}
d_\mu 
& = & d_A {\rm Im} Z_0 +d_B {\rm Im} \tilde{Z}_0 \nonumber \\
& = & \frac{d_A+d_B}{2} ({\rm Im} Z_0 +{\rm Im} \tilde{Z}_0 ) 
     +\frac{d_A-d_B}{2} ({\rm Im} Z_0 -{\rm Im} \tilde{Z}_0 ), 
\end{eqnarray}
where $d_A$ and $d_B$ are the total coefficients of ${\rm Im} Z_0$ 
and ${\rm Im} \tilde{Z}_0$.
Applying unitarity constraints we can define 
the maximal muon EDM \cite{Muon} ($|d_\mu|_{\rm MAX}$) as 
\begin{equation}
|d_\mu|_{\rm MAX}
=  \frac{|d_A+d_B|}{4} \cot\beta ( 1+\tan^2\beta )^{1/2}
  +\frac{|d_A-d_B|}{4} \tan\beta ( 1+\cot^2\beta )^{1/2}. 
\end{equation}
In Figure 3, we present the maximal value allowed by unitarity 
for the muon and the electron EDMs, 
as a function of $\tan\beta$ with $m_0 = 100$, 200 and 400 GeV. 
Also shown is the experimental upper limit 
for the electron EDM ($|d_e|_{\rm U.L.}$). 

In this model, for $\tan\beta \agt 10$, 
$(m_\mu/m_e)|d_e|_{\rm U.L.}$ could be treated 
as an upper limit for the muon EDM since $d_\mu \approx (m_\mu/m_e) d_e$. 
A measured value of the muon EDM above this bound could arise
if the muon EDM and the electron EDM are generated from different sources. 

\begin{figure}
\centering
\leavevmode
\psfig{figure=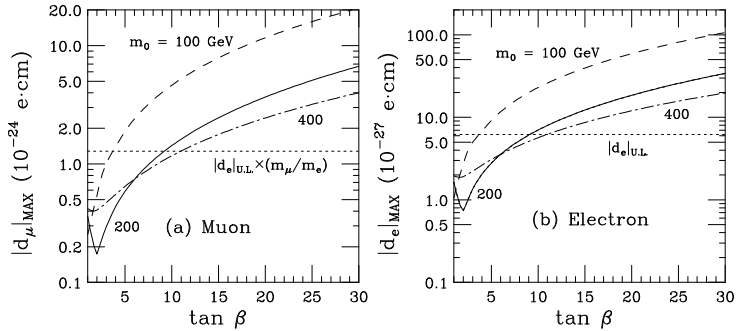,angle=90,height=1.7in}
\caption{
The maximal EDM allowed by unitarity for (a) the muon and (b) the electron. 
\label{fig:muon3}}
\end{figure}

\section{Conclusions}

Our results may be summarized as follows \cite{Muon}:
\begin{enumerate}
\item For $\tan\beta \sim 1$, 
contributions from two loop diagrams involving the $t$ quark 
and the $W$ boson dominate; while for $\tan\beta \agt 10$, 
contributions from two loop diagrams involving the $b$ quark 
and the $\tau$ lepton are dominant. 
\item For $\tan\beta \sim 1$ or $\tan\beta \agt 10$, 
$d_\mu \approx (m_\mu/m_e) d_e$. 
\item For $8 \agt \tan\beta \agt 4$, significant cancellations occur 
among the contributions from two loop diagrams 
and the one loop contribution dominates for $\tan\beta \sim 7$. 
\item For $\tan\beta > 15$ and $m_0 < 300$ GeV, 
Higgs-boson exchange could produce a muon EDM 
which is close to the reach of the proposed BNL experiment.
\end{enumerate}
A positive result for the muon EDM measurement 
could shed light on non-standard CP violation.
A negative result, however, could mean 
\begin{enumerate}
\item that $\tan\beta$ is small than 10; or, 
\item the CP violation parameters ${\rm Im}Z_0$, and ${\rm Im}\tilde{Z}_i$ 
are smaller than their unitarity bounds; or, 
\item the masses of the neutral Higgs scalars and the Higgs pseudoscalar are 
very close to one another. \cite{Weinberg}
\end{enumerate}

\section*{Acknowledgments}
This research was supported in part by the U.S. Department of Energy 
under Grants 
No. DE-FG05-87ER40319 (Rochester) and 
No. DE-FG02-95ER40896 (Wisconsin),
and in part by the University of Wisconsin Research Committee 
with funds granted by the Wisconsin Alumni Research Foundation.



\end{document}